\documentclass[a4paper,11pt]{article}\pdfoutput=1
\usepackage{jheppub} 

\usepackage{graphicx}
\usepackage{amsmath}
\usepackage{amssymb}
\usepackage{rotating} 
\usepackage{xspace}
\usepackage{color}

\usepackage{placeins}
\usepackage{rotating}
\usepackage{cancel}

\newcommand{\likeJ}{\mathcal{L}_{\rm Joint}}
\newcommand{\like}{\mathcal{L}}
\newcommand{\Ohsq}{\Omega_\chi h^2}
\newcommand{\BR}{BR}
\newcommand\RBtaunu{\frac{\BR(B_u \to \tau \nu)}{\BR(B_u \to \tau \nu)_{SM}}}

\newcommand\brbsmumu{\BR(\overline{B}_s\to\mu^+\mu^-)}
\newcommand\brbdmumu{\BR(\overline{B}_d\to\mu^+\mu^-)}

\newcommand{\brbsgamma}{BR(\bar{B} \rightarrow X_s\gamma) }

\def\LSP {$\chi^0_1$}
\def\stau{$\tilde{\tau}_1$}

\def\stop{$\tilde{t}_1$ }

\title{Dark Matter, Sparticle Spectroscopy and Muon $(g-2)$ in $SU(4)_c \times SU(2)_L \times SU(2)_R$}

\author[a]{M. E. G\'omez,}
\author[b,c]{S. Lola,}
\author[d]{R. Ruiz de Austri,}
\author[e]{Q. Shafi}

\affiliation[a]{Departamento de F\'isica Aplicada, Facultad de Ciencias Experimentales, Universidad de Huelva, 21071 Huelva, Spain}

\affiliation[b]{Institute of Nuclear and Particle Physics, NCSR ‘Demokritos’, Agia Paraskevi, 15310, Greece}

\affiliation[c]{on leave from Department of Physics, University of Patras, 26500 Patras, Greece}

\affiliation[d]{Instituto de F\'isica Corpuscular, IFIC-UV/CSIC, Valencia, Spain}

\affiliation[e]{Bartol Research Institute, Department of Physics and Astronomy, University of Delaware, Newark, DE 19716, USA}

\emailAdd{mario.gomez@dfa.uhu.es}
\emailAdd{magda@physics.upatras.gr}
\emailAdd{rruiz@ific.uv.es}
\emailAdd{shafi@bartol.udel.edu}

\abstract{We explore the sparticle mass spectra including LSP dark matter within the framework of supersymmetric $SU(4)_c \times SU(2)_L \times SU(2)_R$ (422) models, taking into account the constraints from extensive LHC and cold dark matter searches. The soft supersymmetry-breaking parameters at $M_{GUT}$ can be non-universal, but consistent with the 422 symmetry. We identify a variety of coannihilation scenarios compatible with LSP dark matter, and study the implications for future supersymmetry searches and the ongoing muon g-2 experiment.}

\begin{document} 
\maketitle
\flushbottom



\section{Introduction}

In recent years, a large body of experimental data, including Higgs
boson measurements \cite{higgs1,higgs2} and cosmological observations
\cite{WMAP1,WMAP2,Ade:2013zuv,Ade:2015xua}, have provided increasingly strong constraints
on new physics beyond the Standard Model (SM). Nonetheless, some new
physics is required to explain, for instance, the observed solar and
atmospheric neutrino oscillations, provide a plausible dark matter (DM) 
candidate, explain the observed baryon asymmetry in the universe, 
help understand electric charge quantization, etc. 

Among the many plausible SM extensions, supersymmetric theories have
several theoretical advantages, including a compelling explanation of
the origin of DM through the lightest supersymmetric
particle (LSP) \cite{DM-susy1,DM-susy2}, and amelioration of the well-known 
fine tuning problem. Despite the fairly strong LHC \cite{CMSdat, ATLASdat} 
and DM \cite{LUX, Akerib:2018lyp, Aprile:2015uzo, XENON1T,Aalbers:2016jon,PICO} constraints on supersymmetry (SUSY), there still remain several viable 
possibilities \cite{SUSYunif,SUSYunif1, Athron:2017qdc,Athron:2017yua,Bagnaschi:2016afc,Fowlie:2013oua,vanBeekveld:2016hug,Cabrera:2016wwr}.
In this paper we investigate a class of supersymmetric models based on
the gauge symmetry $SU(4)_c \times SU(2)_L \times SU(2)_R$ (422) \cite{PS,lr,PS2}, which
have several interesting features. Electric charge
quantization is built in, neutrinos have non-zero masses via the
see-saw mechanism, and the observed baryon asymmetry can be explained
via leptogenesis. Furthermore, the MSSM $\mu$ problem is readily
resolved \cite{mu-PS} in 422, and inflation can also be nicely implemented 
\cite{inflation-PS}.

Because of its gauge structure, the 422 model naturally allows one to
consider non-universal soft SUSY breaking masses at $M_{GUT}$
for the gluino and scalar sectors, leading to significant differences
from other GUTs (Grand Unified Theories). Note also that left-right symmetry 
may not hold at $M_{GUT}$. We explore the implications for particle
spectroscopy focusing, mostly, on the yet to be found supersymmetric
partners of the SM particles, as well as LSP DM. We identify
a variety of coannihilation scenarios that are compatible with the
current searches at the LHC and the presence of 
primordial LSP DM. In addition, supersymmetric contributions to the 
anomalous magnetic moment of the muon $(g-2)$ could help explain the discrepancy 
between the SM prediction and the experimental value \cite{Davier:2010nc}.
In particular, we identify models and SUSY mass relations for which the 
neutralino relic density is consistent with the cosmological bounds and 
explore how their parameter space is constrained by the LHC data.
These predictions will be tested by the ongoing and future DM and LHC searches. 

\section{The $SU(4)_c \times SU(2)_L \times SU(2)_R$ model} 
\label{sec:3}

We start by briefly reviewing the salient features of the 422 model 
\cite{PS,PS2}, which shares many features, but also shows fundamental differences from
standard GUTs such as $SU(5)$ and $SO(10)$. The 422 gauge 
symmetry can be obtained from a spontaneous breaking of
SO(10) by utilizing either the 54 dimensional or the 210 dimensional
representation. The breaking of SO(10) with a Higgs 54-plet yields two
connected components, namely the 422 subgroup and 
$\Sigma_{67} . 422$, where 
$\Sigma_{67}$ is a rotation by $\pi$ in the 6-7 plane 
\cite{KLS-PRD, LS-PLB159}. 

Instead of $\Sigma_{67}$, we could alternatively use the rotation C given by
$C=(\Sigma_{23}) (\Sigma_{67})$,
which is also an element of $SO(10)$. 
This C-transformation interchanges the left-handed and right-handed 
fields and conjugates the representations.
The $SO(10)$ breaking with a Higgs 210-plet also yields the 422 symmetry, 
but the C-symmetry (and left-right (LR) symmetry) is explicitly broken
in this case.

Previous investigations of particle spectroscopy in 422 models have
relied on the presence of left-right symmetry \cite{PS-09}, in order to keep the
number of soft SUSY breaking parameters to a minimum. In this
paper, we go a step further and assume that the soft scalar masses do
not necessarily respect the discrete left-right symmetry. 
In principle, in the left-right asymmetric 422 model, the soft gaugino
masses are not necessarily equal, $M_{SU(2)L}\neq M_{SU(2)R}$, and the SM hypercharge
generator is given by
\begin{equation}
M_1=\frac{3}{5} M_{2R} + \frac{2} {5} M_4,
\label{eq:M4}
\end{equation}
where the $SU(4)$ gaugino mass parameter $M_4$ will be identified with
$M_3$.
Then, if the 422 gaugino masses are independent, this will also hold for the SM
gaugino masses. 
With additional  assumptions, the number of free parameters can be reduced.  
Here, we will  follow the approach of Ref.~\cite{lr,PS-09}  for the gaugino
sector. Supplementing 422 with a discrete left-right C-symmetry, reduces the number 
of independent gaugino masses from three to two. Indeed, while the gaugino masses 
associated with $SU(2)_L$ and $SU(2)_R$ are the same, the gluino mass, 
associated with $SU(4)_c$, in principle can be different. 
The hypercharge generator from
422 implies:
\begin{equation}
M_1=\frac{3}{5} M_2 + \frac{2} {5} M_3 .
\label{eq:M1}
\end{equation}

Our framework is the following: we assume that SUSY breaking 
occurs in a hidden sector at a scale $M_X > M_{GUT}$, via a mechanism that 
generates flavour-blind soft terms in our visible sector. Between the scales 
$M_X$ and $M_{GUT}$, while the theory still preserves the 422 symmetry,
renormalisation and additional flavour symmetries
may induce non-universalities for soft terms that belong to
different representations (while particles that belong to 
the same representation have common soft masses).

We employ GUT relations among the soft terms derived from the
unification group structure \cite{EGLR,Okada:2013ija,
  coannih,Kowalska:2015zja,Kowalska:2014hza,Ellis:2016tjc}. 
 The soft terms for the scalar fields in an irreducible representation 
$r$ of the 422 unification group are defined as multiples of a 
common scale $m_0$: 
\begin{equation}
m_{r}=x_r \, m_{0}, 
\end{equation}
while the trilinear terms are defined as
\begin{equation}
A_r = Y_r \,  A_0,  \;\;\; A_0=a_0 \, m_0 .
\end{equation}
Here, $Y_r$ is the Yukawa coupling associated with the $r$
representation and $a_0$ is a dimensionless factor, which is representation
independent (the representation dependence is taken into account 
in the Yukawa couplings).

In view of the above discussion, we expect the following:
\begin{itemize}
\item Gluino masses: 
We assume the relation in eq.~(\ref{eq:M1}) among gaugino masses. 
We will see that this relation will yield gluino coannihilation as a
viable scenario
\cite{gluino-Co1,gluino-Co2}, 
which was absent in other groups, namely SO(10), SU(5) and flipped SU(5) \cite{EGLR}.
\item 
Soft masses: Sfermions are accommodated in 16-dimensional spinor
representations and their soft mass parameter is $m_{16}$. The electroweak MSSM doublets
lie in the 10-dimensional representation with D-term contributions
that result in splitting of their soft masses. 
Indeed, $m^2_{H_{u,d}} = m^2_{10} \pm 2M_D^2$, and, in our notation: 
\begin{equation}
x_u=\frac{m_{H_{u}}}{m_{16}}, \;\;\;\;\;\;\;\;\;\;
x_u=\frac{m_{H_{d}}}{m_{16}}, \;\;\;\;\;\;\;\;\;\;
\end{equation}
with  $x_u<x_d$.
\item 
LR asymmetric 422: 
In this case there is additional freedom, as the left-right 
asymmetry introduces a new parameter
\begin{equation}
x_{\,_{LR}}=\frac{m_L}{m_R}, \;\;\;\;\;\;\;\;\;\;
\end{equation}
where $m_L$ is the mass of the left-handed sfermions (that preserve
the definition of  $m_{16}=m_0$), and $m_R$ the mass of the corresponding 
right-handed ones.
\end{itemize}

\section{Exploring the model: Methodology}
\label{sec:method}

We perform parameter space scans similar to \cite{EGLR}, where the initial 
conditions of the soft terms are determined by a unification group that breaks 
at $M_{GUT}$ (defined as the scale where the $g_1$ and $g_2$ couplings meet, 
while $g_3(M_{GUT})$ is obtained by requiring $\alpha_s(M_Z)=0.187$).  
For our analysis we use Superbayes
\cite{Bertone:2011nj,Strege:2012bt,Bertone:2015tza}, 
a package to perform 
statistical inference of SUSY models which is linked to SoftSusy \cite{softsusy} 
to compute the SUSY spectrum, to MicrOMEGAs \cite{MicrOMEGAs} and DarkSUSY \cite{DarkSUSY} 
to compute DM observables, SuperIso \cite{SuperIso}
to compute flavour physics and the muon $g-2$, and it uses 
Multinest \cite{multinest} for sampling the parameter space of the models.

The likelihood function, which drives our exploration of regions of the parameter 
space where the model predictions fit the data well, is built from the following 
components:
\begin{equation}
\begin{aligned}
\ln \likeJ &= \ln \like_{\rm EW}+ \ln \like_{\rm B(D)} + \ln \like_{\Ohsq} \\
&+ \ln \like_{\rm DD} + \ln \like_{\rm Higgs} + \ln \like_{\rm SUSY} + \ln \like_{\rm g-2}.
\end{aligned}
\label{eq:like}
\end{equation}
Here:
\begin{itemize}
\item
$\like_\text{EW}$ is the part corresponding to electroweak precision
observables, where constraints from LEP and Tevatron are implemented
as summarised in \cite{ALEPH:2005ab, PDG}.
\item
$\like_\text{B}$ stands for B-physics constraints, from
$\brbsgamma$, $R_{\Delta M_{B_s}}$, $\RBtaunu$,
$\brbsmumu$ and $\brbdmumu$, assuming Gaussian likelihoods
\cite{Strege:2014ija}. For $\brbsmumu$ and $\brbdmumu$ we quote the total uncertainties found by adding in quadrature the theoretical~\cite{Arbey:2012ax} and experimental~\cite{CMSandLHCbCollaborations:2013pla,CMS:2014xfa} uncertainties.
\item
$\like_{\Ohsq}$ is for measurements of the cosmological DM relic
density. Assuming that the lightest neutralino is the dominant DM
component, we include constraints from Planck temperature and lensing
data $\Ohsq = 0.1186 \pm 0.0031$ \cite{Ade:2015xua}, with a (fixed)
theoretical uncertainty $\tau = 0.012$, following Refs. \cite{Strege:2012bt,Roszkowski:2009sm,Roszkowski:2014wqa}, to account for numerical uncertainties.
\item
$\like_\text{DD}$ is for constraints from direct DM detection
searches; we apply data from the Xenon-1T \cite{XENON1T} and PICO-60 \cite{PICO} experiments. The likelihood is computed with the {\tt DDCalc} code~\cite{Savage:2015xta}, and for the computation of the spin-independent and spin-dependent neutralino-nucleon cross-sections, we adopt hadronic matrix elements determined by lattice QCD \cite{QCDSF:2011aa,Junnarkar:2013ac}.
\item
$\like_\text{Higgs}$ implements bounds obtained from Higgs searches
at LEP, Tevatron and LHC via {\tt HiggsBounds} \cite{Bechtle:2013wla}
and LHC Higgs-boson bounds  \cite{higgs1, higgs2}. For this we use {\tt HiggsSignals} \cite{Bechtle:2013xfa}, assuming a 2 GeV theoretical uncertainty in the lightest Higgs mass computation.
\item
$\like_\text{SUSY}$ stands for sparticle searches at colliders. The constraints from SUSY searches at LEP and Tevatron are evaluated following the prescription proposed in~\cite{deAustri:2006pe}. 
\item
$\like_{\text g-2}$: We use the value $\delta a_{\mu}^{\text{SUSY}} =
(28.7 \pm 8.2) \times 10^{-9}$ \cite{Davier:2010nc}, which corresponds
to a $3.6 \sigma$ discrepancy with the SM prediction and relies on $e^+e^-$ data.
\end{itemize}

The {\tt MultiNest} \cite{multinest} algorithm is used to scan the
parameter space and identify regions compatible with the data, though we do not perform 
any statistical interpretation of the results. Instead, we select only
model points that predict the value of all observables within the
2$\sigma$ interval (with $\sigma$ obtained by summing in quadrature
the experimental and theoretical errors); however, we go to 3$\sigma$
for muon $g-2$. We combine the samples produced using logarithmic and linear priors of the model parameters. We finally produce scatter plots showing the correlations of pairs of parameters and/or observables in various planes.


\section{Results of the parameter space scan}
\label{sec:results}

As mentioned above, we perform two scans: one with logarithmic priors 
that scan over a wide range of parameters as shown in Eq. \ref{ranges} and another 
one with flat priors, that are appropriate for looking for correlations. In 
the first case, we find many points with Higgsino DM and resonances
in the annihilation channels, while flat priors are more appropriate 
when searching for coannihilations. 

The 422 non universal soft masses are parametrized using the following
definitions: 
\begin{eqnarray}
100 \text{GeV}\leq& m_0=m_L& \leq 10 \text{TeV} \nonumber\\
-3000 \text{GeV}\leq& M_{3}&\leq 5 \text{TeV} \nonumber\\
50 \text{GeV}\leq& M_{2}&\leq 5 \text{TeV} \nonumber\\
-10  \text{TeV}\leq& A_{0}&\leq 10 \text{TeV} \nonumber\\
2\leq& \tan\beta &\leq 65 \nonumber\\
-1\leq&  x_u&\leq 2 \nonumber\\
0\leq&  x_d&\leq 3 \nonumber\\
-3\leq & x_{LR}  =m_R/m_L  & \leq 3. 
\label{ranges}
\end{eqnarray}
Here $M_1$ is determined by eq.~(\ref{eq:M1}). Note that $M_3$ and $m_R$ are allowed to be negative.

It is well known that if the required amount of relic DM is
provided by neutralinos, particular mass relations must be present in the 
supersymmetric spectrum \cite{Ellis:1999mm, Gomez:1999dk,Boehm:1999bj, Mizuta:1992qp, Edsjo:1997bg, Profumo:2004wk, Feldman:2009zc, Chattopadhyay:2005mv}. We therefore use these mass relations,
together with the neutralino composition, in order
to classify the points that pass the constraints discussed in Sec. 2, according to the following criteria:
\paragraph{Higgsino \LSP  :}
\begin{equation}
h_f >0.1, \;\;|m_A-2 m_\chi| > 0.1 \, m_\chi.
\label{criterio_higgsino}
\end{equation}
The Higgsino-like fraction of the lightest neutralino mass eigenstate is characterized by
the quantity
\begin{equation}
h_f \; \equiv \; |N_{13}|^2 + |N_{14}|^2 \, ,
\end{equation}
where the $N_{ij}$ are the elements of the unitary 
mixing matrix that correspond to the Higgsino mass states.
In this case, 
the lightest chargino $\chi^\pm_1$ and the second lightest neutralino $\chi^0_2$ 
are almost degenerate in mass with the \LSP.
The couplings to the SM gauge bosons are not suppressed and \LSP\ pairs have large 
cross sections for annihilation into $W^+ W^-$ and $ZZ$ pairs, which may reproduce the observed value
of the relic abundance. Clearly, coannihilation channels involving $\chi^\pm_1$ and 
$\chi^0_2$ also contribute. 
\paragraph{$A/H$ resonances:}
\begin{flalign}
|m_A-2 m_\chi|\leq 0.1 \, m_\chi.
\label{criterio_res}
\end{flalign}
The correct value of the relic abundance is achieved thanks to $s$-channel annihilation,
enhanced by the resonant $A$ propagator. The thermal average $\langle \sigma_{ann}v\rangle$
spreads out over the  peak in the cross section, so that neutralino masses for 
which $2m_{\chi} \simeq m_A$ is not exactly realized can also experience resonant annihilations.
\paragraph{$\tilde{\tau}$ coannihilations:}
\begin{flalign}h_f <0.1,\;\;(m_{\tilde{\tau}_1}-m_\chi)\leq 0.1 \, m_\chi.
\label{criterio_RR}
\end{flalign}
The neutralino is bino-like, annihilations into leptons through $t$-channel slepton exchange 
are suppressed, and coannihilations involving the nearly-degenerate \stau \ are necessary 
to enhance the thermal-averaged effective cross section.
\paragraph{$\tilde{\tau}-\tilde{\nu}_\tau$ coannihilations:}
\begin{flalign}
h_f <0.1,\;\;(m_{\tilde{\tau}_1}-m_\chi)\leq 0.1 \, m_\chi,\;\;(m_{\tilde{\nu}_\tau}-m_\chi) \leq  0.1 \, m_\chi.
\label{criterio_LL}
\end{flalign}
This is similar to the previous case but, in addition, the $\nu_{\tilde \tau}$ 
is nearly degenerate in mass with \stau.
\paragraph{$\tilde{t_1}$ coannihilations:}
\begin{flalign}
h_f <0.1,\;\;(m_{\tilde{t}_1}-m_\chi)\leq 0.1 \, m_\chi.
\label{criterio_stop}
\end{flalign}
The \stop \ is light and nearly degenerate with the bino-like neutralino.
These coannihilations were found to be present also 
in the flipped SU(5) model, but not in SO(10) or SU(5) \cite{EGLR}.

What is particularly interesting in the 422 model, which distinguishes it
from the other GUT groups, is that in this case
we get {\em three additional modes of coannihilation}, namely: 
\begin{itemize}
\item {\bf $\tilde{\chi}^+$ coannihilations:}
\begin{flalign}
h_f <0.1,\;\;(m_{\tilde{\chi}^+}-m_\chi)\leq 0.1 \, m_\chi.
\label{criterio_stop}
\end{flalign}
The lightest chargino  is light  and nearly degenerate with the bino-like neutralino.
\item{\bf $\tilde{g}$ coannihilations:}
\begin{flalign}
h_f <0.1,\;\;(m_{\tilde{g}}-m_\chi)\leq 0.1 \, m_\chi,
\label{criterio_stop}
\end{flalign}
since the gluino can be relatively light and nearly degenerate with
the bino-like neutralino. 

\item{\bf $\tilde{b}$ coannihilations:}
\begin{flalign}
h_f <0.1,\;\;(m_{\tilde{b}}-m_\chi)\leq 0.1 \, m_\chi,
\label{criterio_stop}
\end{flalign}
since, due to the LR asymmetry, the $\tilde{b}$  can be light  and
nearly degenerate with the bino-like neutralino \cite{sbottom-Co}.
\end{itemize}
%

\subsection{GUT inputs and Planck compatible regions}

In this subsection we present the phenomenological consequences of
relaxing the universality of the SUSY breaking terms following the 422
pattern. Specifically, we concentrate on the differences with
respect  to the groups based on SO(10) and SU(5) that assume gaugino
mass universality.  As discussed in Sec. 2,  following the 422 group
structure, the gaugino masses are not universal at the GUT scale, and
we also assume left-right  asymmetry for the scalar soft masses at the GUT scale.  
In the figures that follow, we show combined points arising from the linear 
and logarithmic sampling of parameters. In both cases, $\tilde{\chi}^+$  coannihations and 
Higgsino DM are the points found most frequently. 

\begin{figure*}[]
\begin{center}
\vspace*{-2cm}
\includegraphics*[scale=0.6]{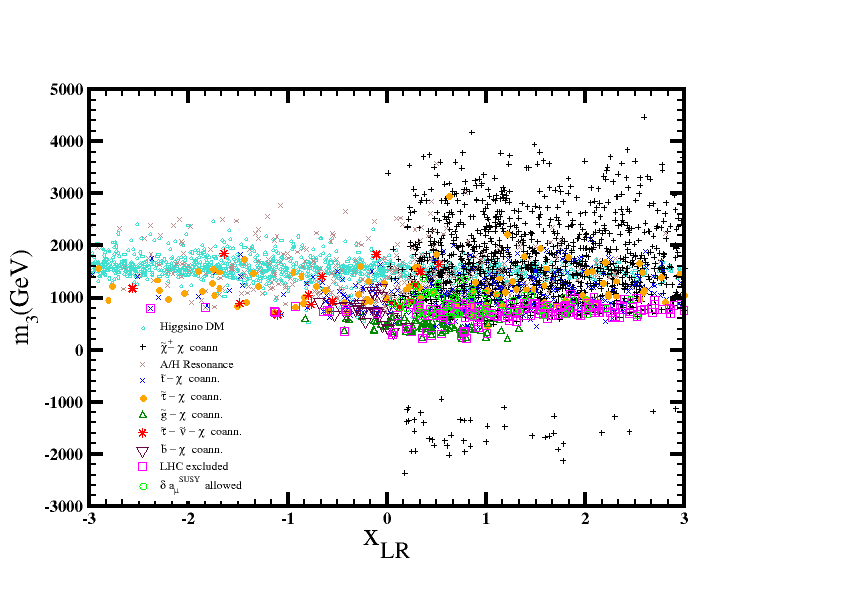}
\vspace*{-1.5cm}
\caption{\it Correlation of the WMAP allowed points with the GUT
  values of $M_3$ and the LR sfermion mass ratio. Different kinds of
  points are denoted with a symbol and color code that also will be maintained in the rest of the plots:  Turquoise dots correspond to Higgsino DM, black crosses to $\chi^\pm- \chi$ coannihilations, brown crosses to A/H resonances, blue crosses to $\tilde{t}- \chi$ coannihilations, orange dots to  $\tilde{\tau}- \chi$ coannihilations, green up triangles to  $\tilde{g}- \chi$ coannihilations, red stars to $\tilde{\tau}- \tilde{\nu}-\chi$, and maroon down triangles to  $\tilde{b}- \chi$ coannihilations.  In addition, green circles enclose points that provide a SUSY contribution to $\delta a_\mu^{SUSY}$ compatible with the experimental bounds, while points enclosed 
  in magenta squares are excluded in our analysis of LHC results (see Sec. 5) .}
\label{fig:m3_xlr}
\end{center}
\end{figure*}
In {\bf Figure}~\ref{fig:m3_xlr} we clearly observe 
that the vast majority of points lie in the upper right region. 
Points with $\tilde{\chi}^+$ coannihilations have a preference for $x_{LR}>0$.  
We find that obtaining the correct prediction of $m_h$ imposes a
correlation between the signs of $M_3$ 
and $A_0$.  The majority of models satisfying this constraint
correspond to $M_3>0$ and $A_0<0$,
however, a few models  with $M_3<0$ and $A_0<0$ are also found.

We observe that most of the classes  of models  satisfying the Planck 
constraints can be found even if sfermion LR symmetry is preserved. However,  points with
$\tilde{\tau}-\tilde{\nu}_\tau-\chi$ (red asterisks)  and
$\tilde{b}-\chi$ (maroon down triangles) coannihilations appear only when the  $LR$
symmetry is broken ($x_{LR}<1$).  Although  the constraints imposed by the  
anomalous magnetic moment of the muon and the LHC searches 
will be discussed in the following sections, we find it illustrative
to anticipate our results in all the plots. Therefore, we enclose in a
green circle the points that explain the discrepancy of the
experimental bound with respect to the SM prediction at the $3-\sigma$
level. Similarly, points excluded by the LHC searches, according to
the analysis presented in Sec. 5, are enclosed in magenta squares. 
 

\begin{figure*}[!htbp]
\begin{center}
\hspace*{-1.5cm}
\includegraphics*[scale=.3]{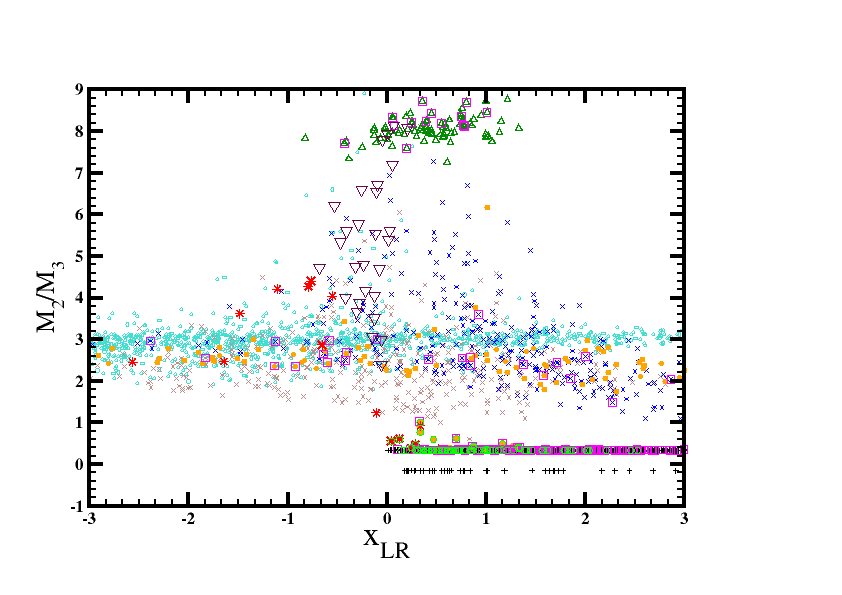}
\hspace{-1.8cm}
\includegraphics*[scale=.3]{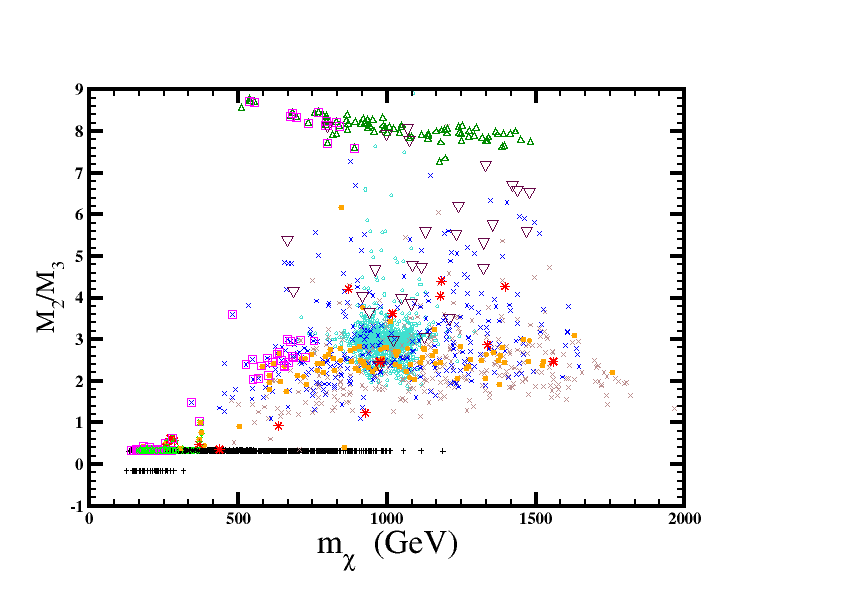}
\hspace{-2cm}
\vspace{-.5cm}
\caption{\it Scatter plots showing the different Planck areas as functions
of the ratios of the GUT values for the soft terms, 
using the same notation as in Figure ~\ref{fig:m3_xlr}}.
\label{fig:m2m3_xlr}
\end{center}
\end{figure*}

The consequences of relaxing the gaugino mass universality can be
appreciated in the left panel of  {\bf Figure~\ref{fig:m2m3_xlr}},
where we can see how eq.~(\ref{eq:M1})  results in narrow ranges of $M_2/M_3$ for which
$m_\chi$ approaches $m_{\tilde{\chi}^+}$; this gives rise to 
$\tilde{\chi}^+-\chi$ coannihilations, almost independently of the neutralino mass and
the LR ratio $x_{LR}$. These coannihilations are associated with 
ratios $M_2/M_3\sim 0.4$ if $M_3>0$, and $M_2/M_3\sim 0.2$ if $M_3<0$.
In both cases, the lightest chargino is mostly wino and $x_{LR}>0$. 
We also find points with $\tilde{\chi}^+-\chi$ coannihilations for
different values of $M_2/M_3$, for cases where $\tilde{\chi}^+$ is not a
pure wino. 
Coannihilations $\tilde{\tau}-\tilde{\nu}-\chi$ where the 
$\tilde{\tau}$ is mostly left-handed, correspond to 
$|x_{LR}|<1$,  while larger values of $|x_{LR}|$ allow
right handed stau dominated  $\tilde{\tau}-\chi$, as well as
$\tilde{t}-\chi$ coannihilations. 
In order to classify coannihilations of the LSP with a sparticle
$\tilde{p}$, we used as a criterion  
a mass ratio $m_{\tilde{p}}/m_\chi=0.1$, although we can see that
coannihilations among several particles are also possible. 
Some coannihilation  points classified as
$\tilde{\tau}-\chi$ 
and $\tilde{\tau}-\tilde{\nu} -\chi$ show  ratios $M_2/M_3$ typical of
the chargino coannihilations, indicating that 
coannihilations $\tilde{\tau}-\tilde{\nu}-\tilde{\chi}^+-\chi$  are possible. 
The neutralino masses satisfying the Planck constraints are displayed in
the right panel of {\bf Figure~\ref{fig:m2m3_xlr}}. 
We can see that points satisfying the muon $(g-2)$ constraints require
values of $m_\chi$  below 500 GeV, and all of them correspond to ratios 
$M_2/M_3$ below 2. Higgsino like neutralino  masses are in the 1 TeV range, 
similar to what was found in other GUTs \cite{EGLR}. We can see
again that  LR asymmetry allows  $\tilde{\tau}-\tilde{\nu}-\chi$ and
$\tilde{b}-\chi$ coannihilations that are not present when the LR symmetry holds ($x_{LR} =1$).

\begin{figure*}[t]
\begin{center}
\vspace{-2cm}
\hspace{-1.5cm}
\includegraphics*[scale=.3]{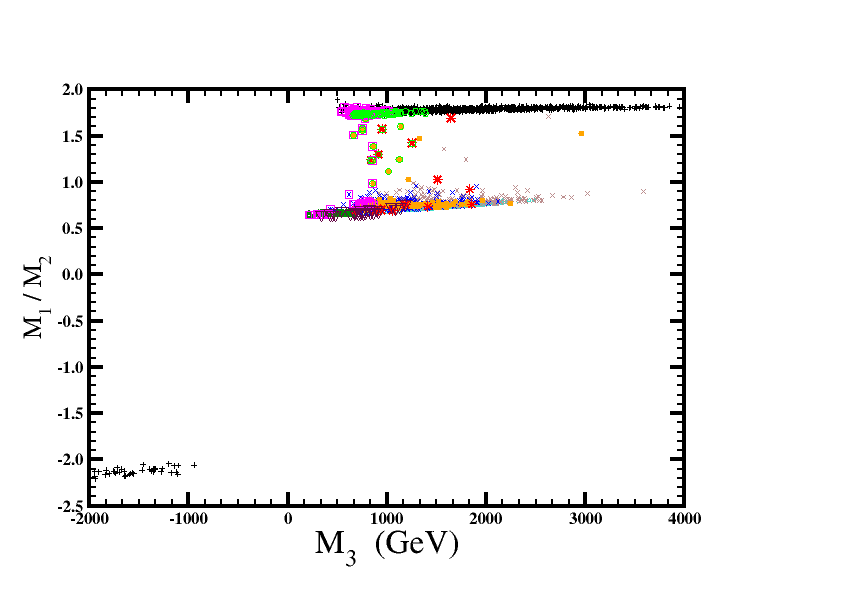}
\hspace{-1.8cm}
\includegraphics*[scale=.3]{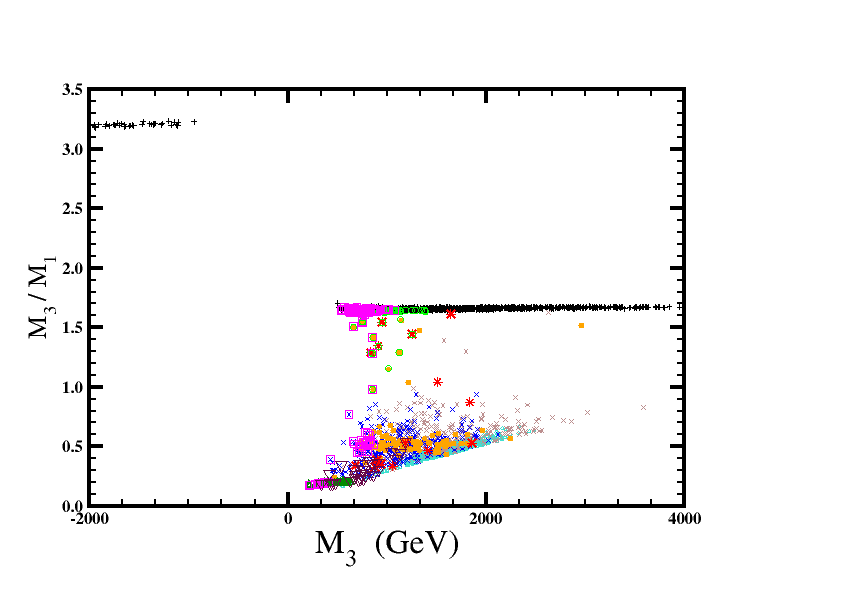}
\hspace{-2cm}
\vspace{-.5cm}
\caption{\it Values of  the gaugino mass ratios vs $M_3$ at the GUT scale
  and their correlation with the different Planck areas as functions of
  the GUT ratios for the soft terms. We use the same notation as in 
Figure \ref{fig:m3_xlr}.}
\label{fig:m3m1_m3}
\end{center}
\end{figure*}

{\bf Figure}~\ref{fig:m3m1_m3} provides a clear picture of the
constraints imposed by gaugino mass relations, as shown in 
eq.~(\ref{eq:M1}), along with the Planck constraints. 
These may imply a second condition on the gaugino masses 
due to relations of the LSP mass with other particles in order to
fulfill the relic density requirements; 
this relation is more diffuse, due to RGE dependence at the low energy
mass scale. 
In the left panel, we can see that two ratios are favored:  
$M_1/M_2\sim 1.8 $ and $M_1/M_2\sim -2.1 $ (the latter corresponds 
to $M_3 < 0$). Due to eq.~(\ref{eq:M1}), these regions can be
correlated to $M_3/M_1\sim 2.2 $ and $M_3/M_1\sim 3.2$, respectively, 
in the right panel. Moreover, $\tilde{g}-\chi$ coannihilations 
are produced by ratios of  $M_3/M_1\sim 0.2 $ that correspond to 
ratios of $M_1/M_2\sim 0.53 $. Points with Higgsino DM and A/H resonances appear 
for narrow ranges of the ratio $M_3/M_1$. This is due to the fact that they impose 
additional constraints on the gaugino masses, decreasing 
the  $\mu$-term so that the Higgsino component of the LSP becomes relevant 
and/or the A/H resonance condition $m_\chi/2\sim m_{A/H}$ is materialised. 
In both cases, the approximate range 0.3-0.5 for $M_3/M_1$ 
is converted by Eq. ~(\ref{eq:M1}) to a range 0.6-0.9 for $M_1/M_2$.  
A very distinct case arises for $\tilde{\chi}^+-\chi$ 
coannihilations, in the case where the neutralino and the chargino are  almost
bino- and wino-like respectively. For sfermion coannihilations, the
relations among the GUT values of the gaugino masses are not as
sharp. Stop coannihilations, due to the effect of the stop mass in the
RGE's, depend on the value of $M_3/M_1$ and, 
through eq.~(\ref{eq:M1}), on $M_1/M_2$ as well.  
Stau coannihilations, however, are not affected by $M_3/M_1$ and can take values 
between the limiting lines characteristic of Higgsino DM and chargino coannihilation.

\begin{figure*}
\begin{center}
\vspace*{-1.5cm}
\hspace*{-1.5cm}
\includegraphics*[scale=0.3]{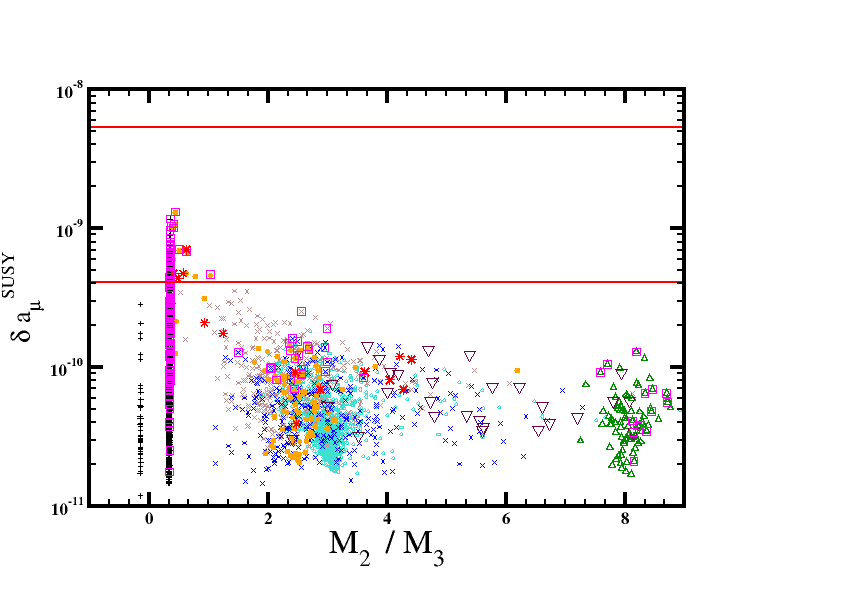}
\hspace*{-1.8cm}
\includegraphics*[scale=0.3]{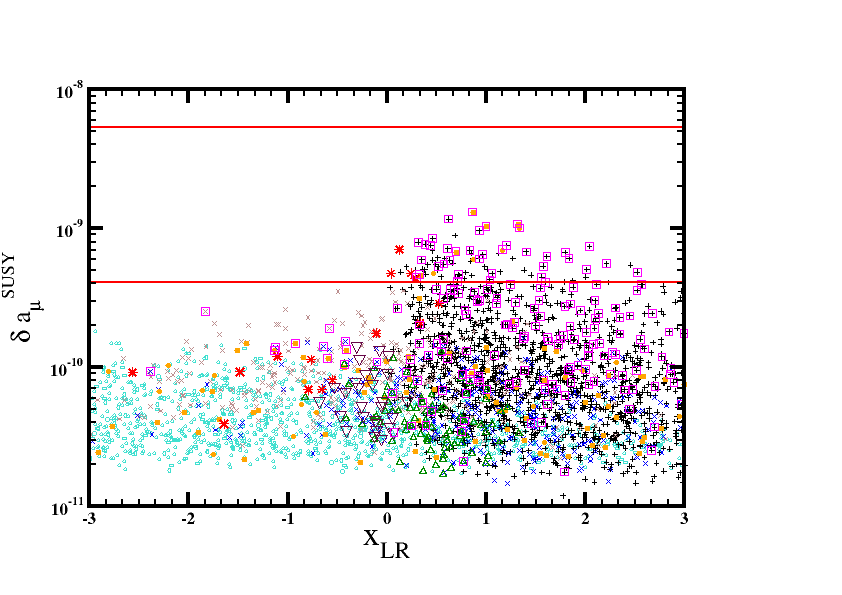}
\hspace*{-2cm}
\end{center}
\begin{center}
\vspace*{-1cm}
\hspace*{-1.5cm}
\includegraphics*[scale=0.3]{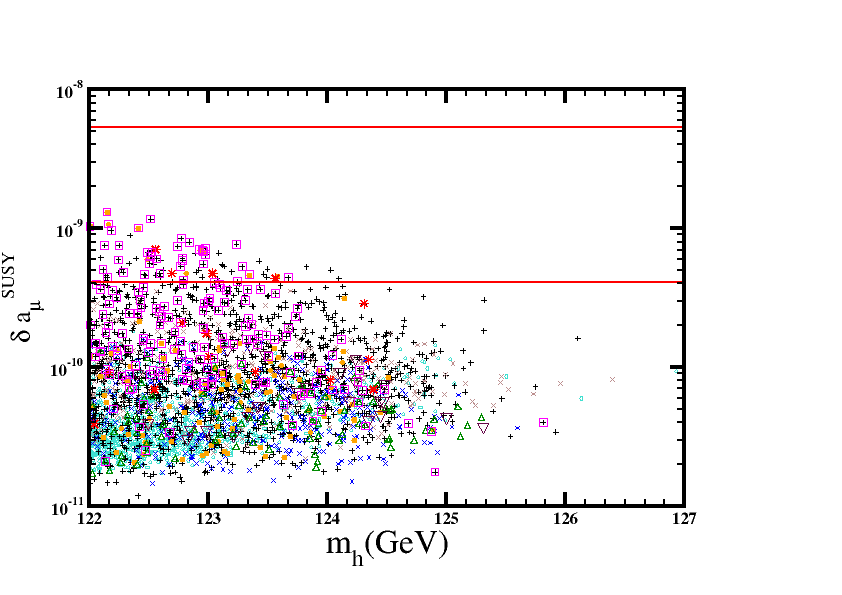}
\hspace*{-1.8cm}
\includegraphics*[scale=0.3]{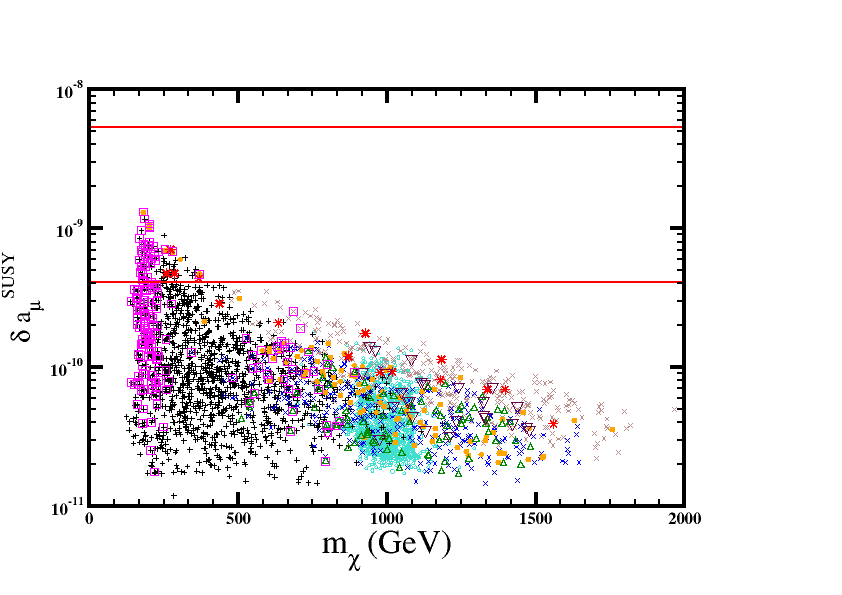}
\hspace*{-2cm}
\vspace*{-.5cm}
\caption{\it The upper panels show the variation of the prediction for  $\delta a_\mu^{SUSY}$ 
with the ratio of the GUT values of $M_2/M_3$ (left) and the LR
asymmetry of the GUT values for the sfermion masses (right). The lower panels
show regions  satisfying $\delta a_\mu^{SUSY}$ and $m_h$
bounds (left) and  $\delta a_\mu^{SUSY}$ vs $m_\chi$ (right). The red
lines are the 3-$\sigma$ bound lines for the experimental discrepancy
of $a_\mu$ with respect to the SM prediction.}
\label{fig:delta_mh}
\end{center}

\end{figure*}

\subsection{Higgs mass and  Muon $(g-2)$}

Connecting the Higgs boson discovery with the lightest neutral SUSY
particle of the MSSM requires a rather heavy SUSY spectrum 
that  makes it
challenging to explain the discrepancy between the experimental value of
$(g_\mu-2)$ and its SM prediction, at least in the simplest SUSY
models. The value  $\delta a_\mu^{SUSY}=(28.7\pm ± 8.2)×10^{-10}$ is
difficult to reach in models with universal soft terms. Even after
allowing non-universalities at the GUT scale for scalar soft terms,
like the models based on SU(5) and SO(10) of Ref.~\cite{EGLR}, the
SUSY contribution remains below the central value.  Therefore, we wish
to investigate whether the pattern of soft terms introduced by 
the 422 symmetry can result in models with a larger contribution to muon $(g-2)$. 
 
To display the relevance of the  particular relation of soft terms
introduced by the 422 symmetry, we present in  the upper panels of 
{\bf Figure~\ref {fig:delta_mh}} the  variation of the prediction of  
$\delta a_\mu^{SUSY}$ with  the GUT gaugino mass ratios (left)  and
the LR asymmetry (right). We can observe that  the highest values of $\delta a_\mu^{SUSY}$ are obtained for $M_2/M_3$ ratios that favor chargino coannihilations. 
We also see that only $M_2/M_3$ below 2 can result in a 
a SUSY contribution compatible with $(g_\mu-2)$.  
The right upper panel shows that the LR soft mass asymmetry
results in some points with $\tilde{\tau}-\chi$ coannihilations 
crossing  the $\delta a_\mu^{SUSY}$ lower bound. These points include cases where the 
stau is mostly left-handed, so that  $\tilde{\tau}-\tilde{\nu}-\chi$ coannihilations 
take place.  

In the lower panels of Figure~\ref {fig:delta_mh},  we explore
the $m_\chi$ values that can simultaneously explain the
experimental value of the Higgs mass and the discrepancy $\delta a_\mu$. 
A SUSY contribution to $\delta a_\mu^{SUSY}$  
above the lower bound is possible for points with chargino
and stau-coannihilations for $m_\chi<500$~GeV. Note  that many of the
points of \cite{Qaisar-am,Gogoladze:2014cha} satisfying $\delta a_\mu^{SUSY}$ are now excluded 
because of the $m_h$ bound (and will be further constrained by the
LHC, as we will discuss in section \ref{sec:lhc}).   

\begin{figure*}[]
\begin{center}
\vspace{-1.5cm}
\hspace*{-1.5cm}
\includegraphics*[scale=.3]{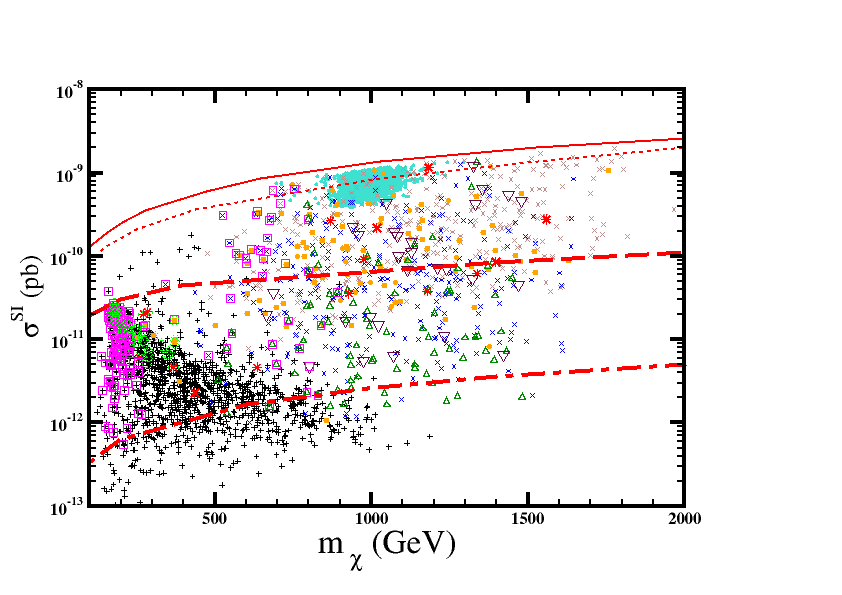}
\hspace*{-1.8cm}
\includegraphics*[scale=.3]{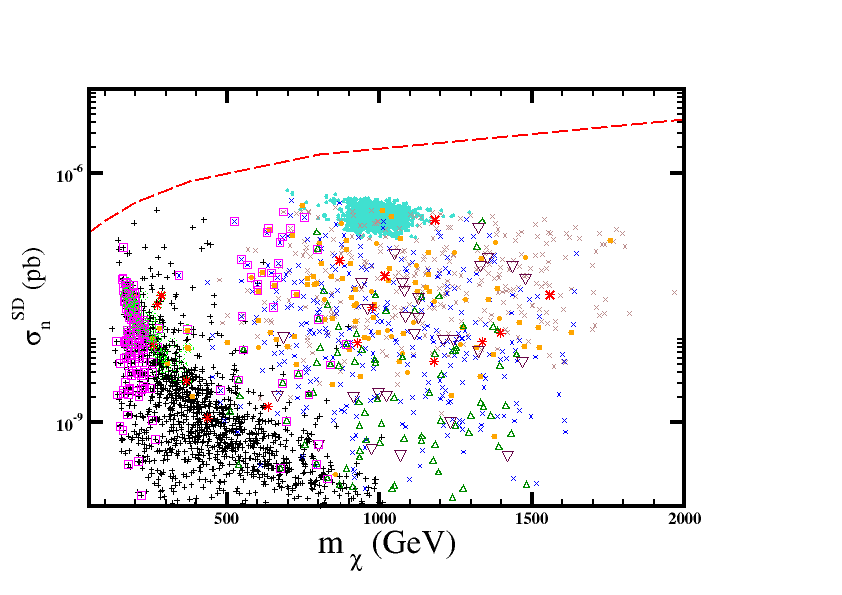}
\hspace*{-2cm}
\vspace*{-.5cm}
\caption{\it Left panel: scatter plot for the SI neutralino-nucleon cross section; the upper red
line corresponds to the Xenon-1T bound \cite{XENON1T} and the dotted line below, to its recent update  \cite{XENON1T_new}.The dash and the dot-dash lines correspond to the projected sensitivities from LZ\cite{Akerib:2018lyp} and DARWIN \cite{Aalbers:2016jon}.
Right panel: same plot for the SD neutralino-neutron cross section and the projected limit from LZ
\cite{Akerib:2018lyp}. }
\label{fig:simlsp}
\end{center}
\end{figure*}

\begin{figure*}[]
\begin{center}
\includegraphics*[scale=.60]{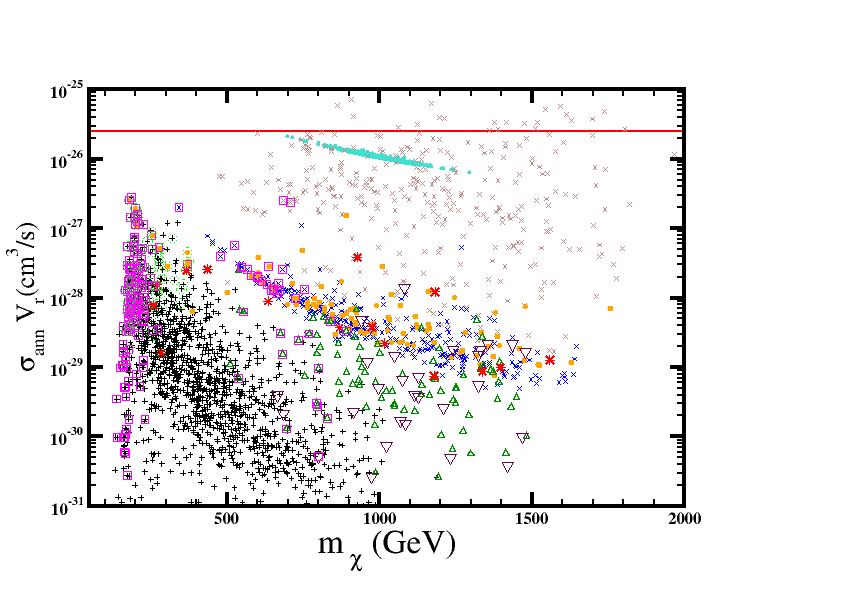}
\vspace{-1cm}
\caption{\it Scatter plot for the total non-relativistic LSP annihilation cross section times 
relative velocity as a function of the neutralino mass. The red line
corresponds to the usual benchmark value of 
$\langle \sigma_{eff} v_\text{rel} \rangle \simeq 2-3 \times 10^{-26}$ cm$^3$/s.}
\label{fig:simlsp2}
\end{center}
\end{figure*}

\subsection{Dark Matter Searches}

The current choice of soft terms allows for models where the
neutralino relic density is located inside the cosmological bounds in
scenarios that imply different relations among SUSY masses. In each
scenario, the composition of the LSP determines its detection prospects.  

In {\bf Figure~\ref{fig:simlsp}} 
we display the Spin Independent (SI) and Spin Dependent (SD) 
neutralino-nucleon cross sections as functions of the neutralino mass,
comparing the theoretical predictions with updated experimental bounds, as summarised in the 
respective figure captions (the line corresponding to the latest announced update from Xenon-1T  \cite{XENON1T_new} is also included). The SI bounds are the most restrictive, and
the current bounds from Xenon-1T \cite{XENON1T} exclude many models where the LSP has a relevant Higgsino component. According to our classification of section 4, these
correspond to Higgsino DM ($h_f>0.1)$. However, points where the 
LSP has a smaller Higgsino component, such as A/H resonances, are on
the scope of coming experiments like LZ
\cite{Akerib:2018lyp}. Furthermore, even models where the LSP
 has a high degree of bino purity can be reached at sensitivities such
 as the ones expected with a multi-ton mass experiment like the DARWIN project~\cite{Aalbers:2016jon}. These experiments can explore most of the models presented here, including  the $(g_\mu-2)$ favored points.

Regarding the SD neutralino-nucleon cross section, bounds from
experiments sensitive to neutralino-proton interactions like
PICO\cite{PICO} are less restrictive than the SI case. Models predicting SI cross sections on the Xenon-1T bound are below the PICO bound by two orders of magnitude.  The predictions for neutralino-neutron cross sections, like the ones displayed in the right panel of {\bf Figure~\ref{fig:simlsp}} are higher. However, we can see on the figure that the LZ prospects still favors the SI over the SD interaction sensitivity. 


{\bf Figure~\ref{fig:simlsp2}} depicts
the total non-relativistic LSP annihilation cross section times 
relative velocity as a function of the neutralino mass. Here as well,
we see that for a subset of points like Higgsino DM and A/H resonances
where the neutralino has an important Higgsino component 
and we hope that further light can be shed in
the near future. 
\begin{figure*}[htbp!]
\vspace{-2cm}
\begin{center}
\includegraphics*[scale=.6]{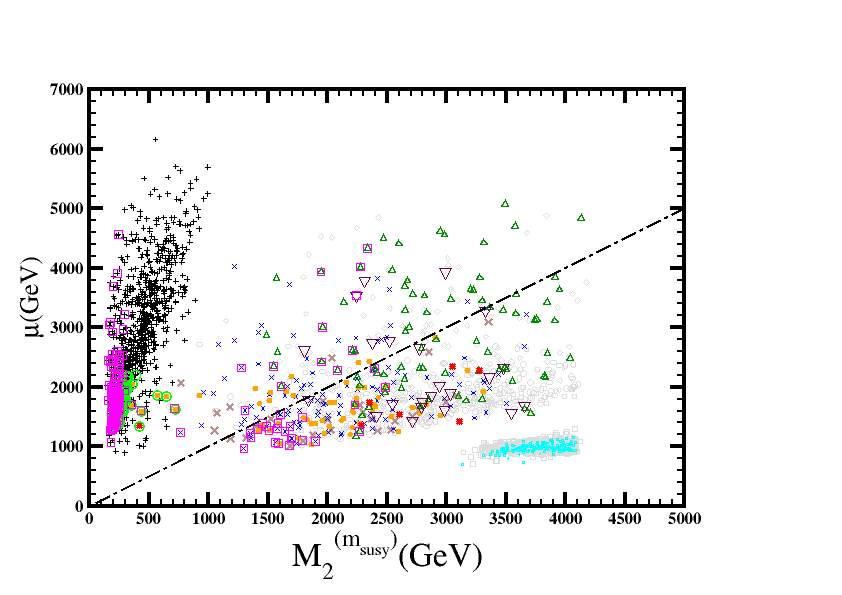}
\vspace{-1cm}
\caption{\it Scatter plots in the $\mu-M_2$ 
plane. Only points analyzed with Smodels are shown with the same legends as in previous 
figures (the brown crosses corresponding to A/H resonances are thicker than in the previous plots). The gray symbols correspond to points not analyzed with Smodels, squares are for Higgsino DM, circles for A/H resonances and diamonds for stop coannihilations. The dot-dash line represents $\mu=M_2$.}
\label{fig:mu-M2}
\end{center}
\end{figure*}
\begin{figure*}
\begin{center}
\vspace*{-1.5cm}
\hspace*{-1.5cm} 
\includegraphics*[scale=0.3]{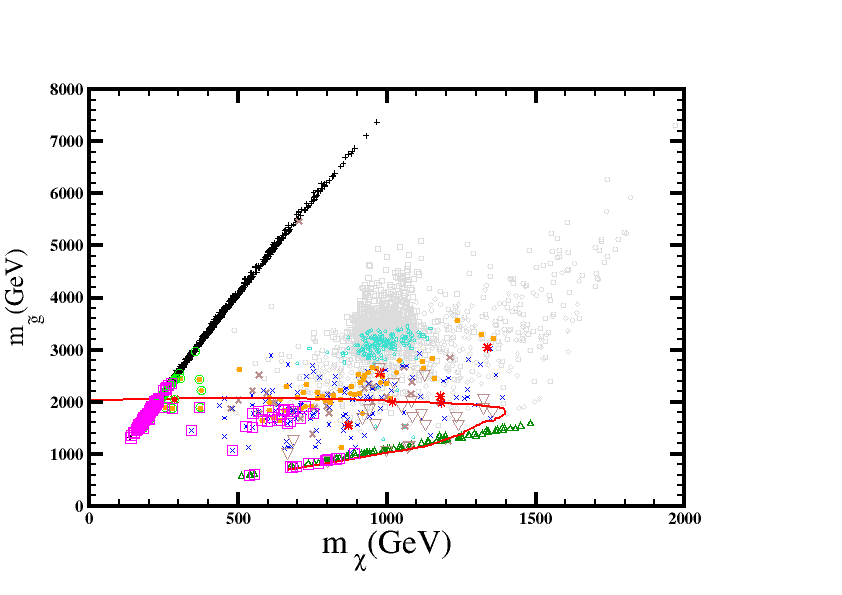}
\hspace*{-1.8cm}      
\includegraphics*[scale=0.3]{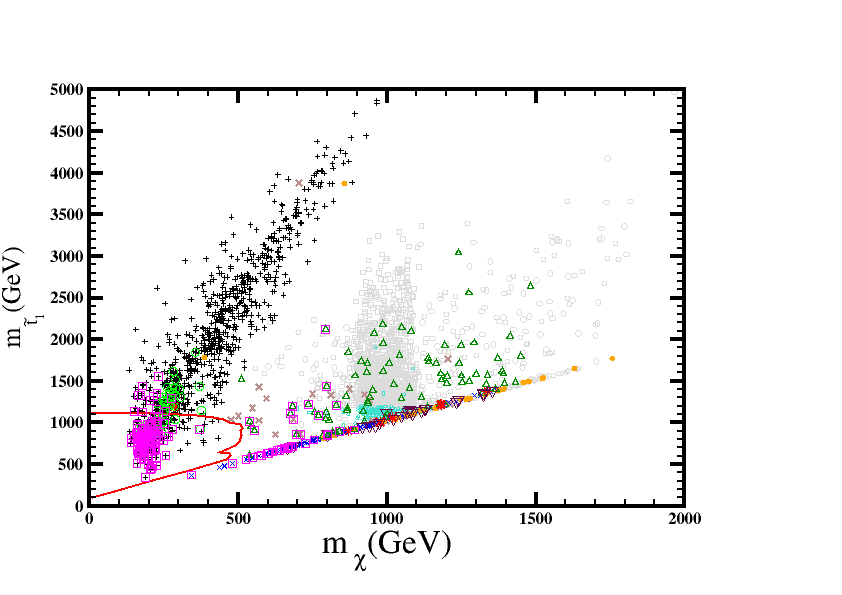}
\hspace*{-2cm}
\end{center}
\begin{center}
\vspace*{-1cm}
\hspace*{-1.5cm}
\includegraphics*[scale=0.3]{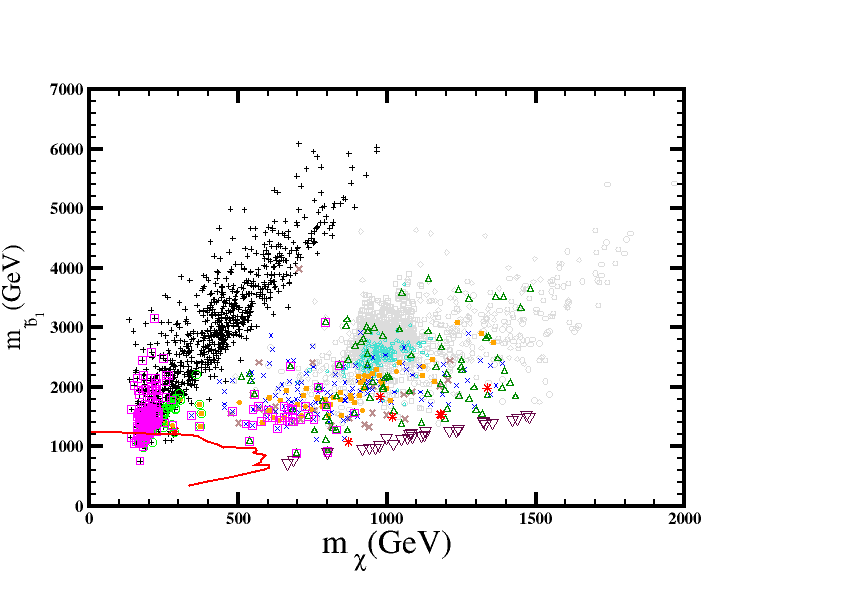}
\hspace*{-1.8cm}
\includegraphics*[scale=0.3]{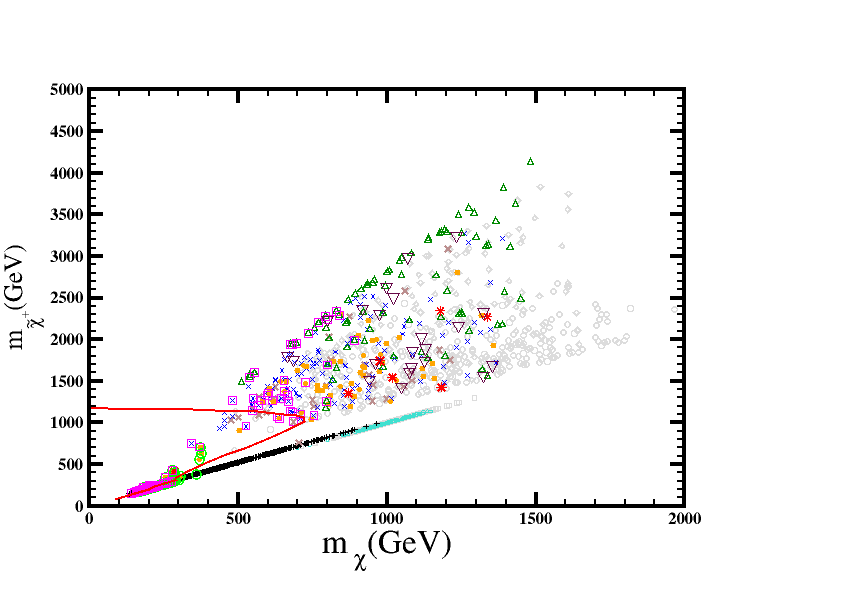}
\hspace*{-2cm}
\vspace*{-.5cm}
\caption{\it The impact of the LHC searches on the 422 models from
  diverse bounds obtained by ATLAS and CMS. For the points we use the
  same code as in Figure ~\ref{fig:m3_xlr}. The solid red line  corresponds to the CMS bound applied to simplified SUSY scenarios while the points in the purple squares are excluded by the same bounds applied to the predictions of our models using Smodels.}
\label{fig:LHC}
\end{center}
\end{figure*}

\section{LHC searches}
\label{sec:lhc}

In previous sections we have seen how the implementation of the 422
group expands the possibilities for DM predictions with respect to
more constrained models. 
In this section, we derive sparticle mass correlations, combining the
experimental and cosmological data summarised in the previous sections
with results from applying the LHC constraints. The results indicate the complementarity of 
DM experiments and of LHC SUSY searches for the asymmetric 422
group, similar to what was found in other GUTs \cite{EGLR}.

The so far unsuccessful searches for SUSY particles impose severe
bounds on their spectrum and interactions. However, it is not
straightforward to translate these bounds to SUSY masses, because the ATLAS 
and CMS experiments typically show results in a model-dependent fashion. 
Namely, the recast of the data is done in the framework of so-called 
Simplified Model Spectra (SMS) that can be considered indicative rather 
than conclusive for real models \cite{SMS_atlas, SMS_cms}. 

Every SMS can be defined by a set of hypothetical particles and a
sequence of their products and decay modes. Therefore, to confront the
theoretical models against the LHC bounds, the predictions must be
expressed in the SMS language. There are several tools designed for
such purpose \cite{ArkaniHamed:2007fw,Drees:2013wra,Smodels} and by using these packages, we can go one step further in comparing our models with LHC data, applying this procedure to a large number of models without the need of a huge computing power \cite{Kraml:2013mwa, Ambrogi:2017lov}. 

In our analysis, we compute for every model its particle mass spectrum using
SoftSusy \cite{softsusy} 
 and the decay branching ratios ($\cal B$) using SUSY-HIT \cite{Djouadi:2006bz}. Then, we pass this information to {\em Smodels-v1.1.1.}~\cite{Smodels} in form of a SLHA \cite{SLHA} file. Production cross-sections ($\sigma$) are calculated by Smodels-v1.1.1 which calls Pythia 8.2 \cite{Sjostrand:2014zea}. 

Smodels-v1.1.1 decomposes production chains in SMS topologies that are confronted with the ones constrained by data. It cannot test all the models we provide, either because their topologies do not match any of the existing experimental results or because their masses fall outside the ranges considered by the experimental searches. These models along with models with weak signals (below 0.05 fb) are considered as beyond the scope of the LHC and classified as not tested. Besides if the mass gap between 
mother and daughter is small, the decays products will be too soft to trigger any signal. We use 5 GeV as the minimum required mass difference for the decay products to be visible. 

In order to present our results, we distinguish among models where the SMS results apply \footnote{SMS results that test the specific topology exist.} and the ones not tested \footnote{When no simplified model result exists for the signal topologies of the point considered.}. For the first, we keep the notation from the previous sections, while not tested models are displayed as gray symbols (squares for Higgsino DM, circles for A/H resonances, diamonds for the $\tilde{t}$-coannihilation). Not tested models are about 50\% of the total investigated models; however this percentage changes depending on the class of models. For clarity reasons we display only the cases where the number of not-tested models dominates over the analysed ones (Higgsino DM, A/H resonances  and $\tilde{t}$-coannihilation). The other classes of models lie in the same areas of the graphs as the displayed points.

In Figure~\ref{fig:mu-M2}, we display $\mu$ versus $M_2$ at
$m_{susy}=\sqrt{m_{\tilde{t}_1} m_{\tilde{t}_2}}$ in order to infer
the composition of the lightest chargino. We can see
that the models with gaugino-dominant charginos are classified as
$\tilde{\chi}^\pm-\chi$ coannihilations; 
we also notice that up to a value of $M_2(m_{susy})\sim 300$~GeV, 
many of these models are affected by the LHC exclusion bounds.

In Figure~\ref{fig:LHC} we show the impact of LHC constraints in some of the mass planes. The results can be summarized as follows: 

\begin{itemize} 
\item On the top left panel we show the impact of strong production through the 0-lepton + jets + $\cancel{\it{E}}_{T}$ channel where the excluded points can be compared with the current coverage by CMS \cite{CMS-SUS-16-033, CMS-SUS-16-036} using SMS results. It is interesting to notice that there are points with gluino masses about 1.3 TeV away from the gluino-compress region, for which the 13 TeV searches should have good sensitivity, which are not excluded by the SMS results. The reason for this is that the produced gluino-pairs decay asymmetrically via, for instance, one into $b\bar{b}$ and the another one into light jets. It is also visible how there are points in the gluino-compress spectra region sensitive to monojet searches. Besides, due to the correlations of the gaugino masses induced by GUT-scale boundary conditions imposed by the model some of the points lying into the chargino-coannihilations region are also excluded by this search. These conclusion hold for the exclusion of squarks. 

\item Next we show the impact of the third generation squarks searches on both stop/sbottom-neutralino mass planes on top right panel and bottom left panel respectively. For stops a reasonable correspondence is found between the sensitivity to the model points and those of the simplified-model decay considered in ATLAS and CMS analyses \cite{ATLAS-SUSY-2013-02, CMS-SUS-16-049,CMS-SUS-16-050,CMS-SUS-16-051}, in which the top squark was assumed to decay to $t + \chi^\pm$. Beyond theses points, points not excluded typically undergo long chain decays. 
For the sbottoms the impact of the LHC constraints is rather weak since for light neutralinos, sbottoms are too heavy to be excluded. Points with sensitivity are well captured by a simplified model where the sbottom decays into a bottom quark and a neutralino. 

\item Finally, the impact of electroweak searches through the multi-leptons + $\cancel{\it{E}}_{T}$ channel \cite{CMS-PAS-SUS-17-004} is shown in the chargino-neutralino mass plane in the bottom right panel. The largest impact of this channel is on the chargino-coannihilation region through the soft two-lepton channel which is sensitive to compress spectrum, specially for wino-like charginos.  As it can be seen, most of the points with $m_{\tilde{\chi^\pm}} \lesssim 300$ GeV are excluded by this search. As it is mentioned this search is complementary to the 0-lepton + jets + $\cancel{\it{E}}_{T}$ search. 

The impact of this search to sleptons is however insignificant because slepton production cross sections are small. The lower values for the slepton masses are of the order of 300 GeV for the staus and 400 GeV for the other generations.

\end{itemize}

In order to compare the 422 LHC predictions with other possible
signals of the 422 modes, we can consider several reference values
for the LSP masses and see the correspondence among different
plots. For instance, 
for values  of $m_\chi$ below 500 GeV we find models that satisfy the
muon (g -2) $3-\sigma$ bound. These points (in green circles) are
classified as models with $\tilde{\chi}+$ or $\tilde{\tau}$
coannihilations. Since their mass spectrum is relatively low, many
of them are excluded by the current LHC bounds according to 
{\em Smodels}. From future
experiments, only the sensitivity of DARWIN will suffice to explore this
area. Their indirect detection signals are also weak since they lie two orders of magnitude below the reference line of Figure~\ref{fig:simlsp2}.

The 422 structure for the soft terms allows coannihilations at values
of $m_\chi\approx 1$~TeV and beyond, which is not possible in models
with universal soft-terms like the CMSSM. The analysis with {\em
  Smodels} indicate that many of the classes of models produce signals
that can be compared with the LHC bounds up to values of  $m_\chi$
about 1.5 TeV. Beyond this mass we find only models with A/H
resonances and $\tilde{t}$ coannihilations that are not tested with
{\em Smodels}. In contrast, the predictions of these models  are the
most promising for indirect detection (see {\bf
  Figure}~\ref{fig:simlsp2}) and they will be tested at Xenon-nT and
LZ, before the sensitivity of DARWIN is reached as we can see in {\bf Figure}~\ref{fig:simlsp2}.

\section{Conclusions}

In this work we explored the predictions of supersymmetric 
$SU(4)_c \times SU(2)_L \times SU(2)_R$ (422) models for supersymmetric particle
spectra, taking into account the constraints from LHC and
cold dark matter searches. The gauge and symmetry breaking
structure of these models leads to very distinct predictions, which
deviate significantly from other models.

In particular, our results are the following:

$\bullet$
A variety of coannihilation scenarios compatible with LSP
dark matter and the LHC have been identified. This clearly
indicates that, despite the fact that no SUSY signal has been found 
so far,  there is still a lot of ground to cover and several
alternative possibilities to explore.

$\bullet$
The particular relations between the gaugino masses in 422 result
in relatively light gluinos with gluino coannihilations, a feature
that is very particular for these models and does not appear in
other GUT schemes. 
Similarly, chargino coannihilations are also found in this case,
and in fact, together with Higgsino DM, are the most
frequently encountered scenarios.

$\bullet$
The fact that the soft supersymmetry 
breaking parameters at $M_{GUT}$ 
can be non-universal, while compatible with the 422 symmetry, gives
rise to additional possibilities and unique features compared
to other GUTs, including stop (also found in flipped SU(5), but not
SO(10) or SU(5))  and sbottom coannihilations.

$\bullet$
We find very concrete predictions for the gaugino
mass ratios that favor scenarios such as 
chargino-neutralino coannihilations. For sfermion-neutralino
coannihilations (and particularly staus where there
is no dependence on $M_3/M_1$), the gaugino mass relations that lead
to viable schemes are not as sharp.

$\bullet$
Overall, the LHC and dark matter searches complement each other in covering the
available parameter space, and among others, accommodate solutions
with prospects for reducing the muon $g-2$ discrepancy via a SUSY
contribution. These solutions are mostly found in the stau-neutralino
and chargino-neutralino coannihilation areas.

\acknowledgments
M.E.G. research was supported by the Spanish MINECO, under grants FPA2014-53631-C-2-P and FPA2017-86380-P. R. RdA is supported by the Ram\'on y Cajal program of the Spanish MICINN, the Elusives European ITN project (H2020-MSCA-ITN-2015//674896-ELUSIVES), the  ``SOM Sabor y origen de la Materia" (PROMETEOII/2014/050) and Centro de excelencia Severo Ochoa Program under grant SEV-2014-0398.
Q.S. acknowledges support by the DOE grant No. DE-SC0013880. We also thank F. M\'arquez Salda\~na for useful discussions.


\end{document}